\documentclass[aps,preprint,superscriptaddress,raggedbottom,nofootinbib]{revtex4-1}
\usepackage{amssymb,amsmath,amsfonts,amsthm,dsfont}
\usepackage{graphicx}
\usepackage[caption=false]{subfig}
\usepackage{float}
\usepackage{lipsum}
\usepackage{array}
\usepackage[normalem]{ulem}
\usepackage{graphicx}
\usepackage{bbm}
\usepackage{physics}
\usepackage{mathtools}
\usepackage{color}
\usepackage{adjustbox}
\usepackage{tabularx}
\usepackage{multirow}
\usepackage{tabu}
\usepackage{makecell}
\setcellgapes{3pt}
\usepackage{algorithm}
\usepackage[noend]{algpseudocode}
\makeatletter
\newcommand{\multiline}[1]{%
  \begin{tabularx}{\dimexpr\linewidth-\ALG@thistlm}[t]{@{}X@{}}
    #1
  \end{tabularx}
}
\makeatother

\newcolumntype{C}[1]{>{\centering\arraybackslash}m{#1}}
\begin{document}
\title{Quantum-classical reinforcement learning for decoding noisy classical parity information}

\author{Daniel K. Park}
\email{dkp.quantum@gmail.com}
\affiliation{School of Electrical Engineering, KAIST, Daejeon, 34141, Republic of Korea}
\affiliation{ITRC of Quantum Computing for AI, KAIST, Daejeon, 34141, Republic of Korea}

\author{Jonghun Park}
\affiliation{School of Electrical Engineering, KAIST, Daejeon, 34141, Republic of Korea}
\affiliation{ITRC of Quantum Computing for AI, KAIST, Daejeon, 34141, Republic of Korea}

\author{June-Koo Kevin Rhee}
\affiliation{School of Electrical Engineering, KAIST, Daejeon, 34141, Republic of Korea}
\affiliation{ITRC of Quantum Computing for AI, KAIST, Daejeon, 34141, Republic of Korea}

\begin{abstract}
Learning a hidden parity function from noisy data, known as learning parity with noise (LPN), is an example of intelligent behavior that aims to generalize a concept based on noisy examples. The solution to LPN immediately leads to decoding a random binary linear code in the presence of classification noise. This problem is thought to be intractable classically, but can be solved efficiently if a quantum oracle can be queried. However, in practice, a learner is more likely to receive data from classical oracles. In this work, we show that a naive application of the quantum LPN algorithm to classical data encoded in an equal superposition state requires an exponential sample complexity. We then propose a quantum-classical reinforcement learning algorithm to solve the LPN problem for data generated by a classical oracle and demonstrate a significant reduction in the sample complexity. Simulations with a hidden bit string of length up to 12 show that the quantum-classical reinforcement learning performs better than known classical algorithms when the sample complexity, run time, and robustness to classical noise are collectively considered. Our algorithm is robust to any noise in the quantum circuit that effectively appears as Pauli errors on the final state.
\end{abstract}

\maketitle
\def\one{{\mathchoice {\rm 1\mskip-4mu l} {\rm 1\mskip-4mu l} {\rm \mskip-4.5mu l} {\rm 1\mskip-5mu l}}}

\section{Introduction}
\label{sec:intro}
Recent discoveries of quantum algorithms for machine learning and artificial intelligence have gained much attention, and stimulated further exploration of quantum technologies for applications in complex data analysis. A type of machine learning considered in this work is related to the ability to construct a general concept based on examples that contain errors. This task can be formulated in the \textit{probably approximately correct} (PAC) framework~\cite{Valiant:1984:TL:1968.1972} in which a learner constructs a hypothesis $h$ with high probability based on a training set of input-output pairs such that $h(x)$ agrees with $f(x)$ on a large fraction of the inputs. In this context, important metrics for characterizing the learnability are the sample complexity and the time complexity that correspond to the minimum number of examples required to reach the goal and the run time of the algorithm, respectively.

A famous example of such tasks is the problem of learning a hidden Boolean function that outputs a binary inner product of an input and a hidden bit string of length $n$ by making queries to an oracle that draws an input uniformly at random. This problem can also be tackled by making queries to a quantum oracle that produces all possible input-output pairs in an equal superposition state. Learning the parity function from a noiseless oracle is easy in both classical and quantum cases.
When the outcomes from an oracle are altered by noise, learning from a classical oracle becomes intractable while the quantum version remains to be efficient~\cite{LPNTheory}. This problem is also known as learning parity with noise (LPN). The LPN problem is equivalent to decoding a random linear code in the presence of noise~\cite{Lyubashevsky2005}, and several cryptographic applications have been suggested based on the hardness of this problem and its generalizations~\cite{Regev:2005:LLE:1060590.1060603,10.1007/978-3-642-27660-6_9}. Furthermore, the robustness of the quantum learning against noise opens up possibilities for achieving a quantum advantage with near-term quantum devices without relying on quantum error correction. However, the existence of a quantum oracle for solving a specific problem is highly hypothetical. In practice, a learner often has to learn from a classical data set. The ability to exhibit the advantage of quantum learning, especially when training examples are classical, remains an interesting and important open problem. 

In this work, we show that a naive application of the quantum LPN algorithm to classical data requires an exponential amount of examples (i.e., training samples) or computing resources, thereby nullifying the quantum advantage. We then propose a quantum-classical hybrid algorithm based on the reinforcement learning framework for solving the LPN problem in the absence of the quantum oracle. The proposed algorithm uses noisy classical samples to prepare an input quantum state that is compatible with the original quantum LPN algorithm. Based on the outcome of the quantum algorithm, a reward is classically evaluated and an action is chosen by a greedy algorithm to update the quantum state in the next learning cycle. Numerical calculations show that the required number of samples and run time can be significantly reduced. Furthermore, simulations show that in the regime of small $n$, the quantum-classical hybrid algorithm performs comparably to or better than the classical algorithm that performs the best in this regime in terms of the sample complexity. Simulation results also suggest that our algorithm is more robust to noise than the classical algorithms. Our algorithm is expected to improve the run time of the classical algorithms, provided that an efficient means to update a quantum state with classical data, such as quantum random access memory, exists. Another notable feature of our algorithm is that it is robust to noise that accumulates to depolarizing error on the outcome prior to the measurement.

Our algorithm also fits within the framework of variational quantum algorithms, a classical-quantum hybrid approach that has been developed as a promising avenue to demonstrate the quantum advantage with noisy intermediate scale quantum (NISQ) devices. Variational quantum algorithms have been used with success to find near-optimal solutions for various important problems in quantum chemistry~\cite{vqe_original,McClean_2016,Romero_2017,ibm_qchem,Moll_2018}, quantum machine learning~\cite{2017arXiv171205771O,PhysRevA.98.032309,2018arXiv180400633S,PhysRevA.98.012324,PhysRevX.8.031084,2019arXiv190400043Z,2019arXiv190100848R}, and quantum control~\cite{PhysRevLett.118.150503}. A unique challenge of the problem considered in this work is that the algorithm must find the \textit{exact} and \textit{unique} solution, i.e., the hidden bit string, with high probability. Thus, our work serves as an intriguing example that utilizes the concept of variational method for finding the exact solution of a problem.

The remainder of the paper is organized as follows. Section~\ref{sec:2} reviews LPN. Section~\ref{sec:3} shows that in the absence of the quantum oracle, the naive application of the quantum algorithm to classical data results in an exponential complexity. In Sec.~\ref{sec:4}, we present a quantum-classical hybrid algorithm based on reinforcement learning for solving the LPN problem. Numerical calculations in Sec.~\ref{sec:4.2} demonstrate that both sample and time complexities of the hybrid algorithm are significantly reduced compared to the native application of the quantum LPN algorithm. Section~\ref{sec:4.3} compares the performance of our algorithm and known classical algorithms via simulations. Section~\ref{sec:5} discusses the resilience to depolarizing errors on the final state, and Section~\ref{sec:6} concludes.

\section{Learning parity with noise}
\label{sec:2}
The goal of the parity learning problem is to find a hidden bit string $s\in\lbrace 0,1\rbrace^n$ by making queries to an example oracle that returns a training data pair that consists of a uniformly random input $x\in\lbrace 0,1\rbrace^n$ and an output of a Boolean function,
\begin{equation}
f(x)=x\cdot s\text{ mod }2.
\end{equation}
A noisy oracle outputs $\left( x, f(x)\oplus e\right)$, where $e\in \lbrace 0,1\rbrace$ has the Bernoulli distribution with parameter $\eta$, i.e., $\mathbbm{P}\left( e=1\right)=\eta$, and $\eta<1/2$~\cite{Angluin1988,Blum2003,Lyubashevsky2005}.
\begin{figure}[t]
\centering
\includegraphics[width=0.7\columnwidth]{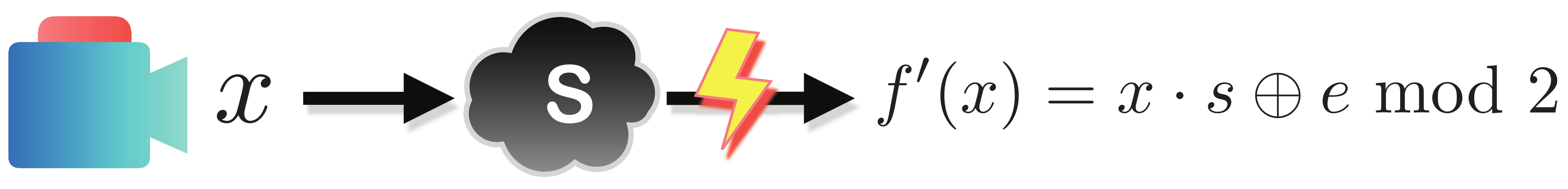}
\caption{\label{fig:LPN_pic}A pictorial representation of the LPN problem.}
\end{figure}

In the quantum LPN algorithm introduced in Ref.~\cite{LPNTheory}, a quantum oracle implements a unitary transformation on the computational basis states and returns the equal superposition of $\ket{x}\ket{f(x)}$ for all possible inputs $x$. By applying Hadamard gates to all qubits at the query output, the learner acquires an entangled state
\begin{equation}
\frac{1}{\sqrt{2}}\left(\ket{0}^{\otimes n}\ket{0}+\ket{s}\ket{1}\right).
\end{equation}
Thus, whenever the label (last) qubit is $1$ (occurs with probability $1/2$), measuring data (first $n$) qubits in their computational bases reveals $s$. Note that this algorithm is very similar to the Bernstein-Vazirani (BV) algorithm~\cite{BernsteinVazirani}, except that in the BV problem the learner can choose an example in each query and the input state of the label qubit is prepared in $\ket{-}$. In the quantum case, since all example data are queried in superposition, the ability to choose an example is irrelevant. On the other hand, the quantum LPN algorithm requires the extra post-selection step since the input of the label qubit is prepared in $\ket{0}$.

\begin{figure}[t]
\centering
\includegraphics[width=0.6\columnwidth]{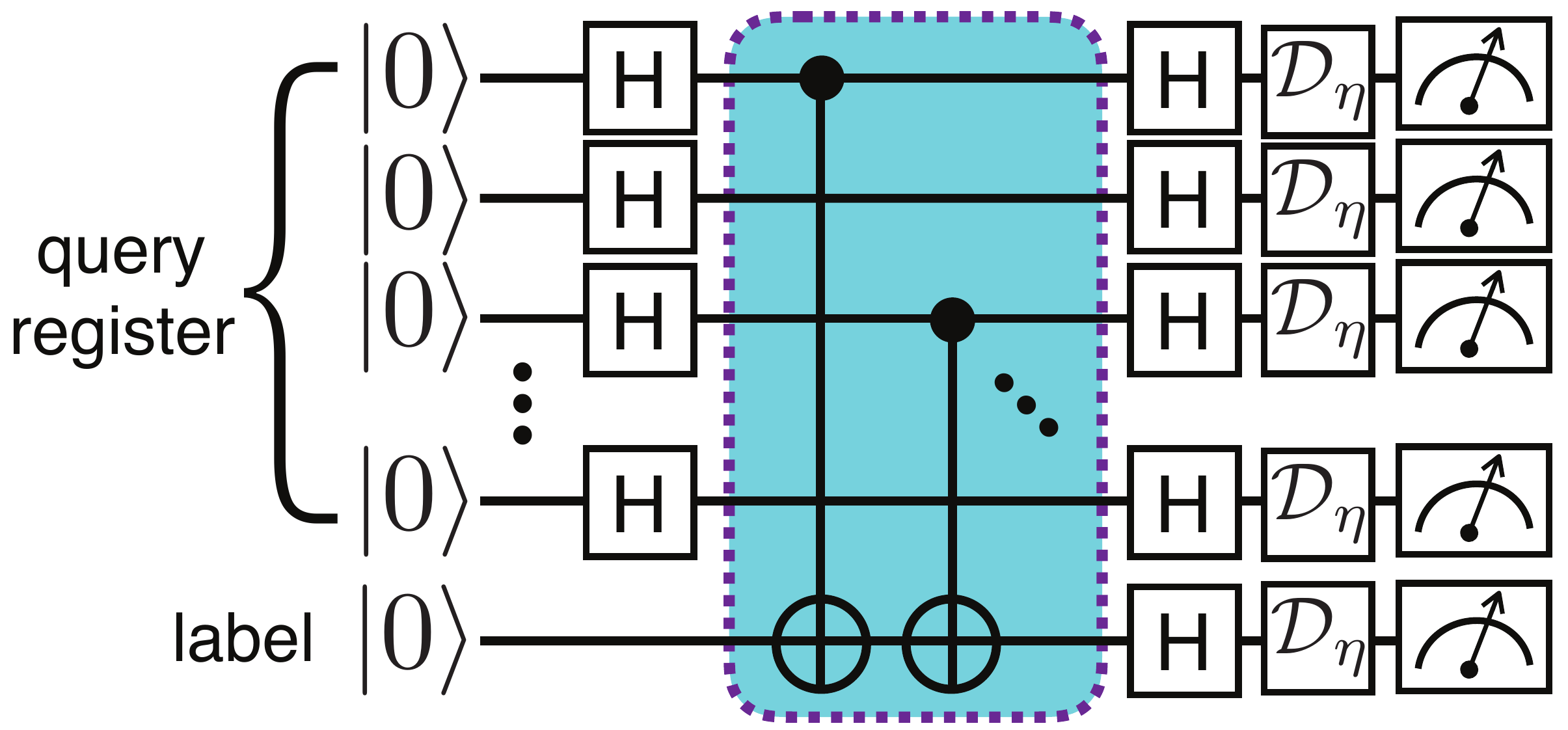}
\caption{\label{fig:LPN0}The quantum circuit for learning parity with noise introduced in Ref.~\cite{LPNTheory}. Hadamard operations ($H$) prepare the equal superposition of all possible input states. The dotted box represents the quantum oracle that encodes the hidden parity function, and is realized using controlled-\texttt{NOT} gates between the query register (control) and label (target) qubits. The hidden bit string in this example is $s=101\ldots 0$. Before measurement, all qubits experience independent depolarizing noise denoted by $\mathcal{D}_\eta$ with a noise rate $\eta<1/2$.}
\end{figure}

A noisy quantum oracle can be modeled with the local depolarizing channel $\mathcal{D}_\eta\left(\rho\right)=\left(1-2\eta\right)\rho+\eta\one$ acting independently on all qubits at the oracle's output with a known constant noise rate of $\eta<1/2$. The quantum circuit for solving the LPN algorithm is depicted in Fig.~\ref{fig:LPN0}. In this example, $s$ is $101\ldots 0$, and it is encoded via a series of controlled-\texttt{NOT} (c-\texttt{NOT}) gates targeting the result qubit controlled by the data qubits. The shaded area in the figure represents the quantum oracle whose structure is hidden from the learner.

Learning the hidden parity function from noiseless examples is efficient for both classical and quantum oracles. However, in the presence of noise, the best-known classical algorithms have superpolynomial complexities~\cite{Angluin1988,Blum2003,Lyubashevsky2005,Levieil2006}, while the quantum learning based on the bit-wise majority vote remains efficient~\cite{LPNTheory}. The query and time complexities of the LPN problem for classical and quantum oracles are summarized in Tab.~\ref{tab:1}
\begin{table}[h]
\centering
\begin{tabular}{c|c|c|C{4.5em}}
Reference & Oracle & Queries (samples) & Time\\ \hline\hline
Angluin \& Laird (AL)~\cite{Angluin1988}& Classical & $O(n)$ & $O\left(2^n\right)$ \\ \hline
Blum, Kalai \& Wasserman (BKW)~\cite{Blum2003} & Classical &$2^{O\left(\frac{n}{\log n}\right)}$ & $2^{O\left(\frac{n}{\log n}\right)}$ \\ \hline
Lyubashevsky (L)~\cite{Lyubashevsky2005}& Classical& $O\left(n^{1+\epsilon}\right)$& $2^{O\left(\frac{n}{\log \log n}\right)}$ \\ \hline
Cross, Smith \& Smolin (CSS)~\cite{LPNTheory} & Quantum & $O\left(\log n\right)$& $O(n)$ 
\end{tabular}
\caption{\label{tab:1} Summary of the query (or sample) and time complexities of various LPN algorithms reported in previous works.}
\end{table}

The advantage of having a quantum oracle for solving an LPN problem was demonstrated experimentally with superconducting qubits in Ref.~\cite{LPNexp}. Furthermore, a quantum advantage can be demonstrated even when all query register qubits are fully depolarized by using deterministic quantum computation with one qubit~\cite{DQC1PhysRevLett.81.5672,PhysRevA.97.032327}.

In the following sections, we discuss quantum techniques to solve the LPN problem in the absence of the quantum oracle. The general strategy considered in this work is to prepare a specific quantum state based on $M$ classical noisy training samples, and apply the measurement scheme developed in Ref.~\cite{LPNTheory}. The measurement outcome of the query qubits in the computational basis, $\tilde{s}_M$, yields the hypothesis function. The goal of our algorithms is to minimize $M$ with which the probability to guess the correct hidden bit string is greater than $2/3$, i.e.,
\begin{equation}
\label{eq:success_condition}
    \mathbbm{P}(\tilde{s}_M=s\vert M)= \gamma,\;\gamma> 2/3.
\end{equation}
Then by repeating the algorithm a constant number of times and taking a majority vote, $s$ can be found with high probability. Note that this is not a strictly necessary condition as the majority vote can find the correct answer efficiently with high probability for $\gamma\geq 1/2+1/\delta$ as long as $\delta$ is at most poly($n$), since the algorithm is to be repeated $O(\delta^{2})$ times. However, Eq.~(\ref{eq:success_condition}) is a sufficient condition to solve the problem.

\section{Naive application of quantum algorithm to classical data}
\label{sec:3}
\subsection{Learning from a sparse set of training samples}
\label{sec:3.1}
Given $M<2^n$ examples of data, $(x_j,f(x_j))$, where $j=1,\ldots,M$, a naive way to apply the quantum LPN algorithm is to create a quantum state,
\begin{equation}
\label{eq:limited_data}
    \ket{\Psi}=\frac{1}{\sqrt{M}}\sum_{j=1}^M\ket{x_j}\ket{f'(x_j)},
\end{equation}
and treat it as the output of the quantum oracle. Then, as in the quantum LPN algorithm, single-qubit Hadamard gates are applied to all qubits and the label qubit is measured in the computational basis. The measurement outcome of $1$, which occurs with the probability of 1/2, is post-selected to leave the query register qubits in the state,
\begin{equation}
\label{eq:limited_sample}
    \ket{\psi_1}=\frac{1}{\sqrt{M2^n}}\sum_{j=1}^M\left\lbrack(-1)^{e_j}\ket{s}+\sum_{y\neq 0^n}(-1)^{x_j\cdot y\oplus e_j}\ket{s\oplus y}\right\rbrack.
\end{equation}
From the above state, the probability to guess $s$ correctly is
\begin{equation}
    \mathbbm{P}(\tilde{s}_M=s)=\frac{M(1-2\eta)^2}{2^n}.
\end{equation}
However, this result also implies that even when the quantum oracle outputs only a fraction of all possible examples as an equal superposition state, and the noise does not act coherently on all $x_j$ as in Refs.~\cite{LPNTheory,LPNexp}, the LPN problem can still be solved.
As a side note, since $x_j$ is drawn uniformly at random, $\mathbbm{P}(x_j\cdot y=1|y\neq 0^n)=\mathbbm{P}(x_j\cdot y=0|y\neq 0^n)=1/2$ and $\mathbbm{E}\left\lbrack\sum_j^M (-1)^{x_j\cdot y}|y\neq 0^n\right\rbrack=0$. Thus, the probability to measure an incorrect bit string $s\oplus y$ (see the second term in Eq.~(\ref{eq:limited_sample})) depends on the error distribution that determines $e_j$. For instance, the worst case occurs when the error bits $e_j$ coincides with $x_j
\cdot y\; \forall j$. In this case, the probability to obtain the incorrect answer is $\mathbbm{P}(\tilde{s}_M=s\oplus y)=M/2^n\ge\mathbbm{P}(\tilde{s}_M=s)$. Thus, in the worst case scenario, the naive application of the quantum algorithm to a limited number of classical samples cannot solve the LPN problem.

\subsection{Data span}
\label{sec:data_span}
In order to reduce the number of queries, linearly independent data can be added to generate an artificial data, which can be used in creating the state in the form of Eq.~(\ref{eq:limited_data}). However, this not only requires the classical pre-processing, but also can increase the error probability of the generated parity bit. For example, from two data pairs, $(x_1,f'(x_1))$ and $(x_2,f'(x_2))$ with linearly independent inputs, a new data $(x_3=x_1\oplus x_2,f(x_3)=f'(x_1)\oplus f'(x_2))$ can be created. Since $f'(x_1)\oplus f'(x_2)=(x_1\oplus x_2)\cdot s\oplus(e_1\oplus e_2)\text{ mod }2$, the error probability of the artificial parity bit is
\begin{equation}
    \eta'=\mathbbm{P}(e_1\oplus e_2=1)=\mathbbm{P}(e_1=1)\mathbbm{P}(e_2=0)+\mathbbm{P}(e_1=0)\mathbbm{P}(e_2=1)=2\eta(1-\eta).
\end{equation}
The above equation can be generalized for a new data generated from $d$ linearly independent data as follows:
\begin{equation}
    \eta'=\mathbbm{P}\left(\sum_{i=1}^{d} e_i\text{ mod 2}=1\right)=\sum_{j=\text{odd}}\binom{d}{j}\eta^j(1-\eta)^{d-j}=\frac{1-(1-2\eta)^d}{2}.
\end{equation}
For $d>1$ and $\eta>1/2$, $\left(1-(1-2\eta)^d\right)/2>\eta$. Therefore, data span always increases the error rate in addition to increasing the time complexity for pre-processing the data. However, when $\eta$ is sufficiently small so that $\eta'$ also remains reasonably small, one may consider using the data span trick in order to reduce the sample complexity.

\subsection{Generation of artificial data with a parity guess function}
\label{sec:GPF}
The next strategy we consider is to generate missing data by guessing the parity function and design an iterative algorithm to improve the accuracy of the guess. In this approach, the rate of accuracy improvement with respect to the number of queries determines the efficiency of the algorithm. 

A brief description of the iterative LPN (I-LPN) algorithm is as follows. First, all $2^n$ examples are provided as an equal quantum superposition state using $M$ real data and $2^n-M$ artificial data generated by a parity guess function. The quantum state can be prepared by guessing the quantum oracle of the quantum LPN algorithm, and inserting the output state of the oracle as an input to quantum random access memory (QRAM)~\cite{QRAMPhysRevLett.100.160501,QRAMPhysRevA.78.052310,PhysRevA.86.010306,1367-2630-17-12-123010,FFQRAM-SciRep-Park} to update its entries according to real data. The circuit-based QRAM introduced in Ref.~\cite{FFQRAM-SciRep-Park} can use \textit{flip-register-flop} processes to update an output of a guessed quantum oracle with real data using the number of steps that increases at least linearly with the number of samples. Then the usual quantum LPN protocol that consists of applying Hadamard gates, projective measurements, and the post-selection outputs an $n$ bit string in the register qubits. This string is used to construct a new parity guess function in the next iteration for which a new sample is also acquired. The learner can also repeat the measurement procedure for guessing a new parity function without querying a new sample. This iteration is referred to as \textit{epoch}. More detailed description of the algorithm is given in steps below.
\begin{algorithm}[H]
\begin{algorithmic}[1]
\vspace{2mm}
 \State Make an initial guess of $s$ as $\tilde{s}_0=0^n$
 \For{$m=1$ {\rm to} $M$}
 \State Collect $(x_m,f(x_m))$
 \If{$x_m=0^n$}
 \State Set $f(x_m)=0$
 \EndIf
\For{$i=1$ to \textit{number of epoch}}
\State \multiline{Use $\tilde{s}_{m-1}$ to implement a quantum oracle and prepare $\frac{1}{\sqrt{2^n}}\sum_j^{2^n}\ket{x_j}\ket{g(x_j)}$, where\\ \vspace{-6mm}$g(x_j)= x_j\cdot \tilde s_{m-1}\text{ mod }2$}
\State Update the above state as $\ket{\Psi}=\frac{1}{\sqrt{2^n}}\left(\sum_{j=1}^{m}\ket{x_j}\ket{f'(x_j)}+\sum_{j=m+1}^{2^n}\ket{x_j}\ket{g(x_j)}\right)$
\State Apply Hadamard gates on all qubits
\State \multiline{Measure the label qubit in the computational basis and post-select the state with the\\ \vspace{-7mm}measurement outcome of $\ket{1}$}
\State Measure the query registers of the post-selected state in the computational basis
\State Set $\tilde{s}_{m-1}$ to the measured bit string
\EndFor
\State Set $\tilde{s}_{m}=\tilde{s}_{m-1}$
\EndFor
  \caption{\label{alg:1}Iterative LPN (I-LPN)}
  \end{algorithmic}
\end{algorithm}

\begin{figure}[t]
\centering
\includegraphics[width=0.7\columnwidth]{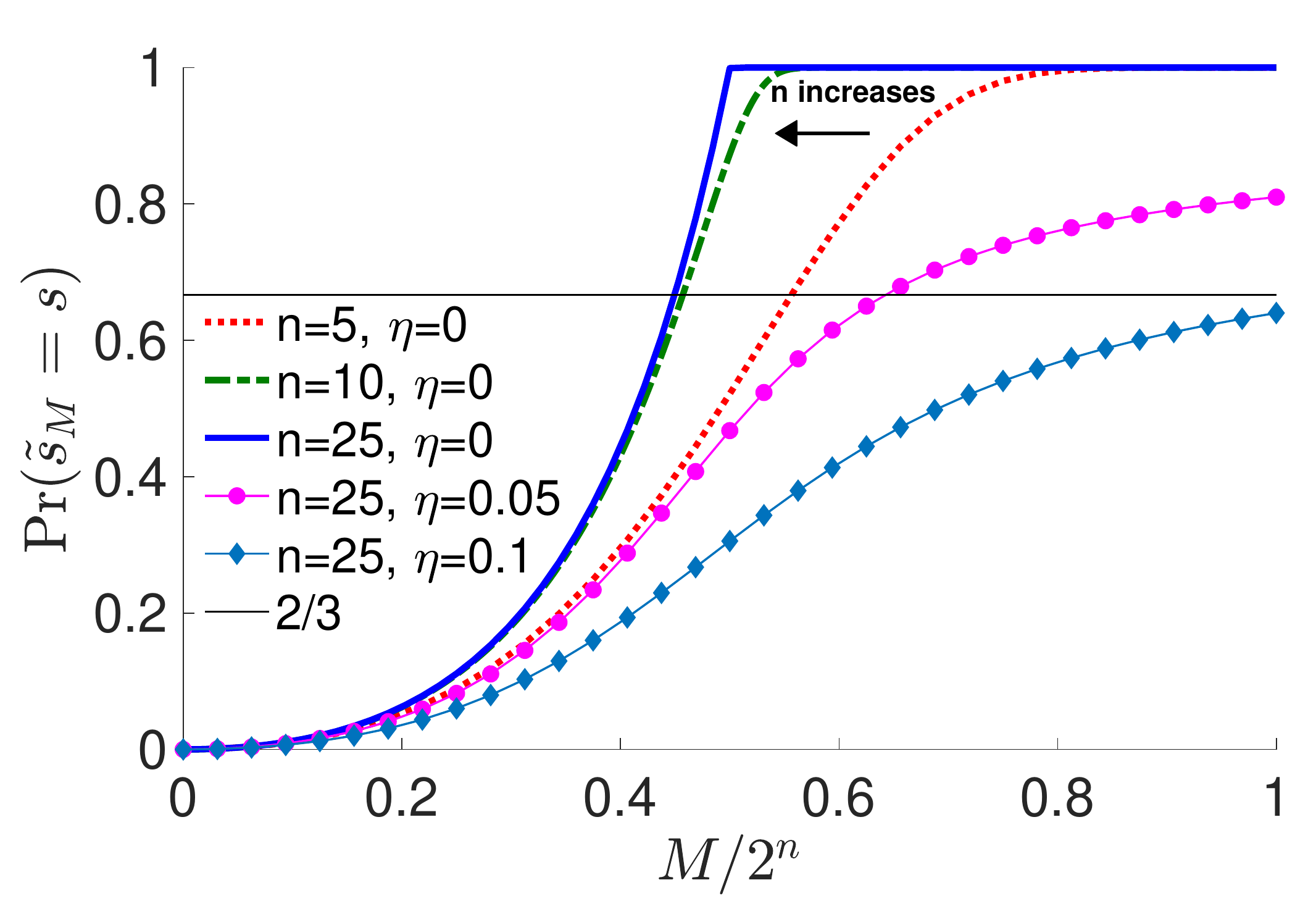}
\caption{\label{fig:naive_update}Success probability of the iterative LPN algorithm with respect to the number of samples in which the parity guess function is obtained from the quantum LPN circuit and used in the subsequent round.}
\end{figure}

Now we analyze the performance of I-LPN. In the iterative algorithm, the time complexity is dominated by the state preparation step. Since the quantum oracle implementation and the QRAM process given $M$ training samples requires $O(n)$ and $O(M)$ run times, respectively, we focus on the estimation of the sample complexity. The I-LPN algorithm uses aforementioned procedure to prepare a quantum state
\begin{equation}
    \ket{\Psi}=\frac{1}{\sqrt{2^n}}\sum_{j=1}^{2^n}\ket{x_j}\ket{h(x_j)},
\end{equation}
where 
\begin{equation}
\label{eq:h1}
h(x_j)=\begin{dcases}
    f'(x_j) = x_j\cdot s\oplus e_j \text{ mod }2\text{ if } j\le M,\\
    g(x_j)  = x_j\cdot \tilde s_{M-1}\text{ mod }2\text{ if } j>M,
\end{dcases}
\end{equation}
and $\tilde s_{M-1}$ is the parity guess function from the previous round. Now let $\varepsilon_j$ be a Bernoulli random variable that is 0 if $h(x_j)=f(x_j)$ and 1 otherwise. In other words, the weight of a string defined as $\varepsilon\coloneqq \varepsilon_1\varepsilon_2\ldots\varepsilon_{2^n}$, denoted by $w(\varepsilon)$, is the number of different bits between $h_{1:2^n}$ and $f_{1:2^n}$, where $\square_{i:j}$ denotes a binary string $\square(x_i)\square(x_{i+1})\ldots \square(x_j)$. With this definition, we can write
\begin{equation}
\label{eq:h2}
    h(x_j)=x_j\cdot s\oplus \varepsilon_j,
\end{equation}
where 
\begin{equation}
\label{eq:vareps}
\varepsilon_j=\begin{dcases}
   e_j\text{ if } j\le M,\\
   x_j\cdot (s\oplus \tilde{s}_{M-1})\text{ if } j>M.
\end{dcases}
\end{equation}
The post-selected state is
\begin{equation}
    \ket{\psi_1}=\frac{1}{2^n}\sum_j^{2^n}\left((-1)^{\varepsilon_j}\ket{s}+\sum_{y\neq 0^n}(-1)^{x_j\cdot y\oplus\varepsilon_j}\ket{s\oplus y}\right),
\end{equation}
The probability to obtain $s$ from the projective measurement in the computational basis is
\begin{equation}
\label{eq:14}
    \mathbbm{P}_M\coloneqq\mathbbm{P}(\tilde s_{M}=s)=(1-2r(M))^2,
\end{equation}
where $r(M)=w(\varepsilon)/2^n$ is the error probability in estimating $f$ from $h$ given $M$ noisy data. From Eq.~(\ref{eq:vareps}), one can see that if $\tilde{s}_{M-1}=s$, then $\varepsilon_j=0$ $\forall j>M$, and only the errors in the real samples yield non-zero values in $\varepsilon$. However, if $\tilde{s}_{M-1}\neq s$, then since $x_j$ is chosen uniformly at random, $\varepsilon_j=0$ for 1/2 of the set of input $x_j$ for $j>M$ on average. With this, the expectation value of the weight of $\varepsilon$ can be calculated as
\begin{equation}
    \mathbbm{E}\left(w(\varepsilon)\right)=\begin{dcases}
    M\eta\text{ if } \tilde{s}_{M-1}=s,\\
    M\eta+(2^n-M)/2\text{ if } \tilde{s}_{M-1}\neq s.
    \end{dcases}
\end{equation}
We start the algorithm with an initial sample $(x_1, f(x_1))$, and the initial guess $\tilde s_0=0^n$. Then the initial error probability is
\begin{equation}
\label{eq:initial_error}
    r(1)=\frac{1}{2^n}\frac{\eta}{2^n}+\left(1-\frac{1}{2^n}\right)\frac{\eta+(2^n-1)/2}{2^n}.
\end{equation}
The first term takes into consideration that $\tilde s_0=0^n$ is the right answer with the probability $1/2^n$. In this case, only the real data can carry incorrect parity bits with a probability $\eta$. The second term indicates that when $\tilde s_0$ is incorrect, 1/2 of the $2^n-1$ guessed parity bits are wrong on average. Using above equations, the error probability in the subsequent rounds up to $M$ samples can be written recursively as
\begin{align}
    r(2)&=\mathbbm{P}_1\cdot \frac{2\eta}{2^n}+\left(1-\mathbbm{P}_1\right)\cdot\frac{2\eta+(2^n-2)/2}{2^n},\nonumber \\
    r(3)&=\mathbbm{P}_2\cdot \frac{3\eta}{2^n}+\left(1-\mathbbm{P}_2\right)\cdot\frac{3\eta+(2^n-3)/2}{2^n},\nonumber \\
    &\vdots \nonumber \\
    r(M)&=\mathbbm{P}_{M-1}\cdot \frac{M\eta}{2^n}+\left(1-\mathbbm{P}_{M-1}\right)\cdot\frac{M\eta+(2^n-M)/2}{2^n}.
\end{align}
For brevity, we denote $\tilde{\eta}_0=M\eta/2^n$ and $\tilde{\eta}_1=\left(M\eta+(2^n-M)/2\right)/2^n$ throughout the manuscript. We plot the success probability as a function of the number of sample for various values of $n$ and $\eta$ in Fig.~\ref{fig:naive_update}. If one increases \textit{epoch}, the fidelity curve simply converges faster to the one for $n\rightarrow \infty$. Thus, the number of samples needed to achieve desired success probability is exponential in $n$.

In the following section, we show that an introduction of a simple policy for updating the parity guess function, as done in reinforcement learning, significantly enhances the learning performance.

\section{Reinforcement learning}
\label{sec:4}
\subsection{Greedy algorithm}
To improve the performance of the iterative algorithm introduced in the previous section, we use the concepts of reinforcement learning, such as state, reward, policy and action. The key addition to the previous iterative algorithm is the use of a greedy algorithm, which always exploits current knowledge to maximize immediate reward, as the policy to make an action. We refer to this algorithm as reinforcement-learning parity with noise (R-LPN).

\begin{figure}[t]
\centering
\includegraphics[width=0.8\columnwidth]{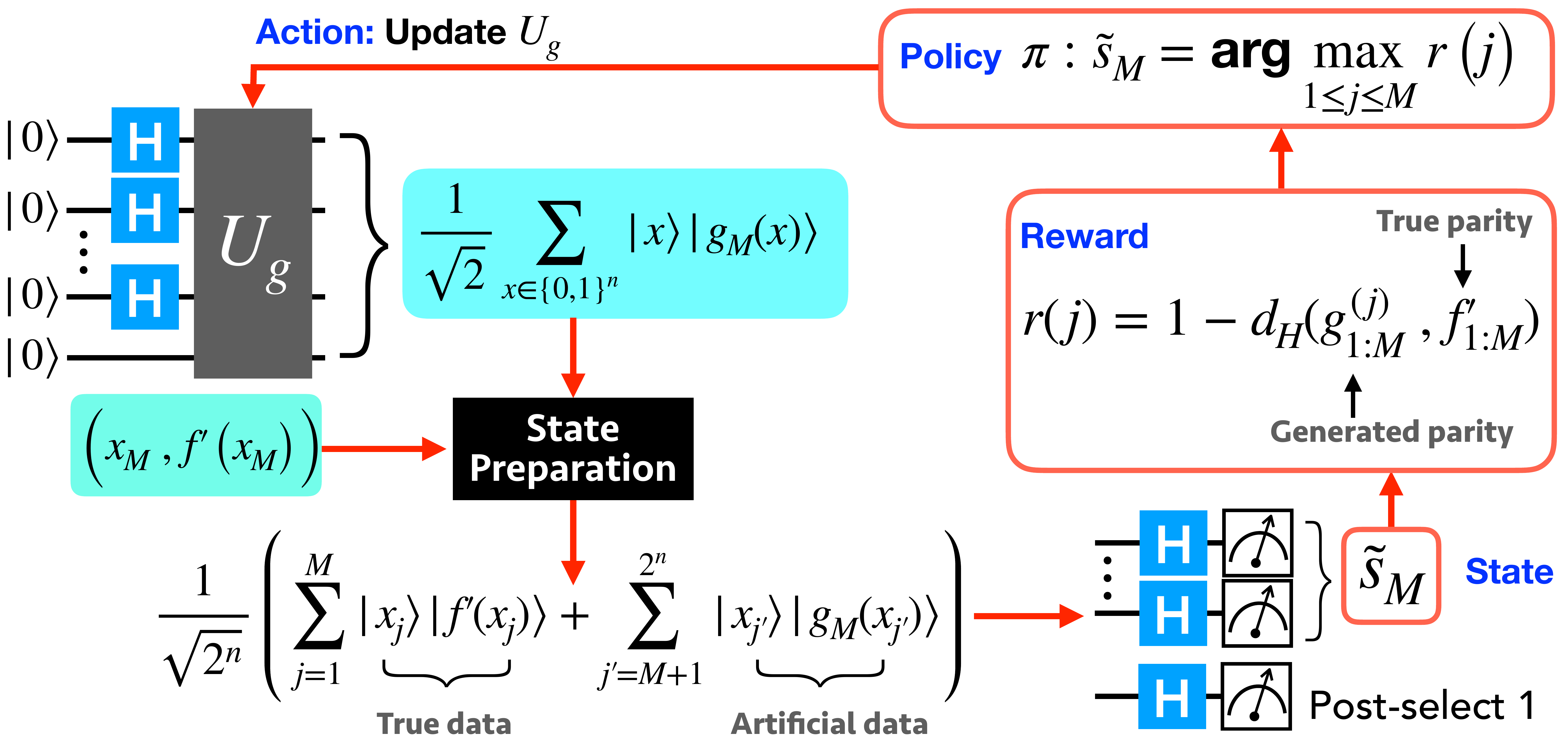}
\caption{\label{fig:R-LPN}Schematic of the quantum-classical hybrid algorithm for solving the learning parity with noise problem, explained by using the terminologies in reinforcement learning.}
\end{figure}

The underlying idea of R-LPN can be described as follows. The state in each iteration is the guessed bit string after performing the usual quantum LPN algorithm. The reward is determined by the Hamming distance between the parity bits generated by the guess and the parity bits of the real data. At $M$th query, the learner obtains $M$ guessed bit strings as well as $M$ reward values. The greedy algorithm then selects the guessed bit string that maximizes the reward, and use it to construct the guessed quantum oracle. Our algorithm can be viewed as a variational quantum algorithm as the guessed quantum oracle can be parameterized with controlled-not gates and is updated in each iteration. The detailed description of the R-LPN algorithm is provided below, and a schematic representation of the algorithm is shown in Fig.~\ref{fig:R-LPN}.
\begin{algorithm}[H]
\begin{algorithmic}[1]
\vspace{2mm}
 \State Make an initial guess of $s$ as $\tilde{s}_0=0^n$
 \For{$m=1 \text{ \rm to }M$}
 \State Collect $(x_m,f(x_m))$
 \If{$x_m=0^n$}
 \State Set $f(x_m)=0$
  \EndIf
 \For{$i=1$ to \textit{number of epoch}}
  \State \multiline{Use $\tilde{s}_{m-1}$ to implement the oracle in the quantum LPN algorithm and prepare $\frac{1}{\sqrt{2^n}}\sum_j^{2^n}\ket{x_j}\ket{g(x_j)}$, where $g(x_j)= x_j\cdot \tilde{s}_{m-1}\text{ mod }2$}
  \State Create a state $\ket{\Psi}=\frac{1}{\sqrt{2^n}}\left(\sum_{j=1}^{m}\ket{x_j}\ket{f'(x_j)}+\sum_{j=m+1}^{2^n}\ket{x_j}\ket{g(x_j)}\right)$
  \State Apply Hadamard gates on all qubits
  \State \multiline{Measure the label qubit in the computational basis and post-select the state with the measurement outcome of $\ket{1}$}
  \State \multiline{Measure the query registers of the post-selected state in the computational basis}
  \State Set $\tilde{s}_{m}$ to the measured bit string
  \State Generate $m$ sets of guessed parity bits $g_{1:m}^{(j)}$, $1\le j\le m$ using $\tilde{s}_{1}$,\ldots, $\tilde{s}_{m}$
  \State Calculate the Hamming distance, $d_H(g^{(j)}_{1:m},f'_{1:m})\; \forall\; 1\le j\le m$
  \State Set $\tilde{s}_{m-1}=\text{arg}\min_{1\le j\le m}d_H(g^{(j)}_{1:m},f'_{1:m})$
  \EndFor
  \State Set $\tilde{s}_{m}=\tilde{s}_{m-1}$
  \EndFor
  \caption{\label{alg:2}Reinforcement LPN (R-LPN)}
  \end{algorithmic}
\end{algorithm}

\subsection{Numerical analysis}
\label{sec:4.2}
We analyze the performance of R-LPN by numerically calculating the  error probability, the probability to measure $\tilde{s}_M\neq s$ in the round with $M$ samples, similar to the recursive calculation shown in Sec.~\ref{sec:GPF}. The algorithm uses parity guess functions from all measurements up to the present round, i.e., $\tilde{s}_1,\ldots,\tilde{s}_M$. To construct the recursive formula, we consider two situations. First, the set of parity guess functions does not contain the answer, i.e., $s\notin\lbrace \tilde{s}_1,\ldots,\tilde{s}_M \rbrace$. This occurs with the probability $p_M=\prod_j^{M-1}\left(1-(1-2r(j))^2\right)$, where $r(j)$ is the error probability at the $j$th round, and $(1-2r(j))^2$ corresponds to the probability to obtain $s$ (see Eq.~(\ref{eq:14})). When the parity guess function is wrong, the probability to measure the wrong hidden bit string in the given round is $\tilde{\eta}_1$ as explained in the previous section. 

When $s\in\lbrace \tilde{s}_1,\ldots,\tilde{s}_M \rbrace$, we further consider two situations. First, $M$ parity bits in the training examples are error-free, which occurs with a probability of $(1-\eta)^M$. In this case, given $M$ linearly independent examples, which can be produced with the probability $\prod_{k=0}^{M-1}(1-2^{k-M})>1/4$ for any $M>1$, there are $\lceil 2^{n-M}\rceil$ choices out of all possible parity guess functions $\tilde{s}_M\in\lbrace 0,1\rbrace^n$ that can generate the same parity bit string as $f_{1:M}$. Note that when $x_j$ is uniformly zero, then $f(x_j)=0$ for any $s$. Thus, we exclude this example when calculating the Hamming distance between the guessed parity bits and the true parity bits. We define $c_M=(\lceil 2^{n-M}\rceil-1)/(2^n-1)$ as the probability to pick a wrong parity guess function that produces the same parity bits as $s$ among $2^n-1$ possible bit strings. Since there are only $M$ parity guess functions, the probability to obtain an incorrect parity function is actually less than $c_M$. However, we use $c_M$ to make a reasonable estimation. The incorrectly guessed parity function produces $(2^n-M)/2$ errors in the artificial parity bits when averaged over uniformly random input.

If the true $M$-bit parity string, $f'_{1:M}$, contains errors, then the probability to obtain a wrong parity guess function can be estimated as $\beta_M=\sum_{k=1}^{\lfloor M\eta\rceil}{M\choose k}c_M$. This means that for simplicity, there are up to $\lfloor M\eta\rceil$ (nearest integer to $M\eta$) errors in the true parity bit string. Combining all cases considered above, the error probability -- the probability to obtain an incorrect parity guess function in the round with $M$ examples -- can be estimated as
\begin{align}
    r(M)=&(1-p_{M-1})\Big{[}\left(1-\eta\right)^Mc_M \left((2^n-M)/2\right)/2^n\nonumber\\
    &+\left(1-(1-\eta)^M\right)\left(\tilde\eta_0(1-\beta_M)+\tilde\eta_1\beta_M\right)\Big{]}\nonumber \\
    &+p_{M-1}\tilde\eta_1,
\end{align}
where the initial error probability, $r(1)$, is given in Eq.~(\ref{eq:initial_error}).

\begin{figure}[t]
\centering
\includegraphics[width=0.7\columnwidth]{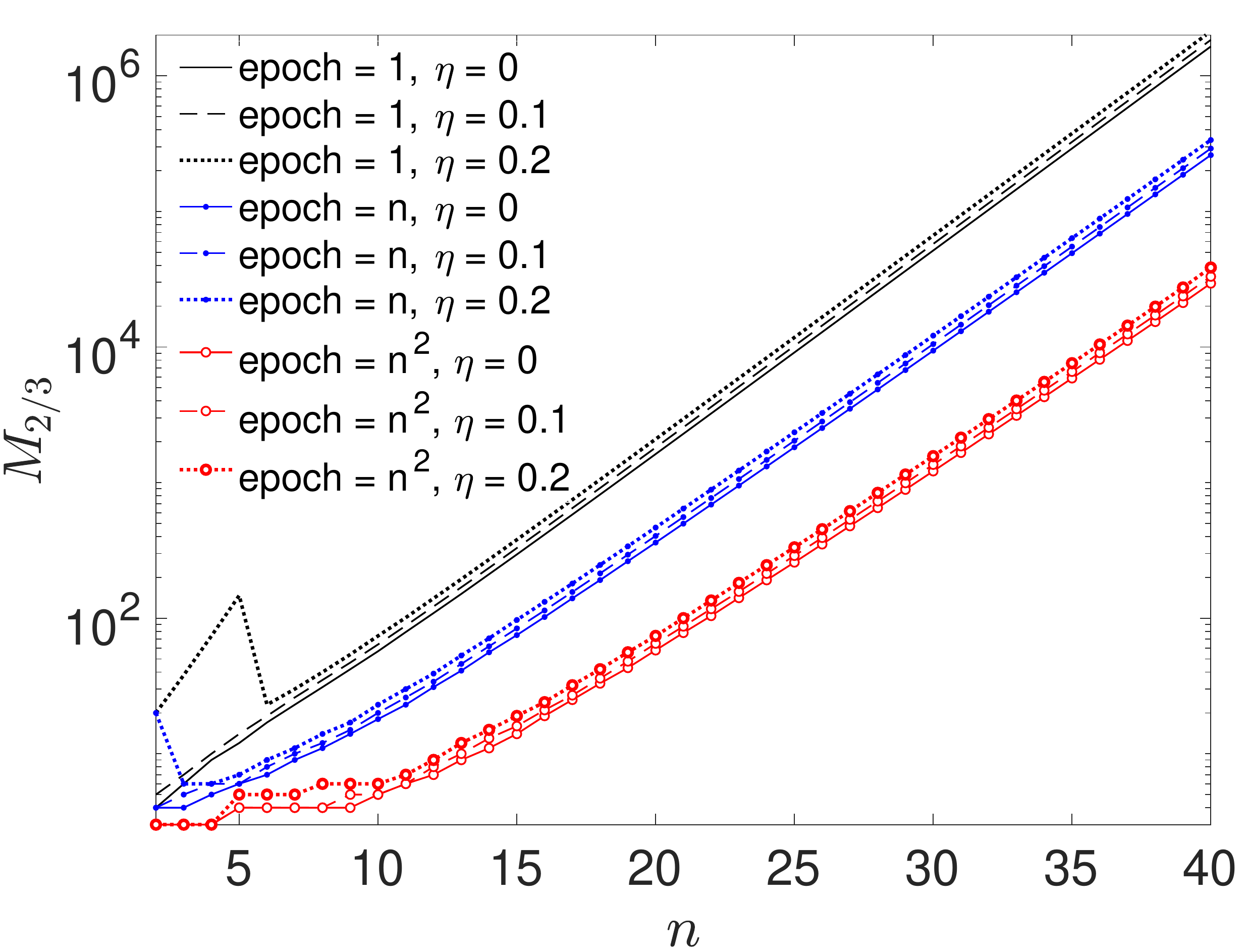}
\caption{\label{fig:numerical}Number of samples required for achieving $\mathbbm{P}(\tilde{s}_M=s)>2/3$, denoted by $M_{2/3}$ as a function of the length of the hidden bit string, $n$, for various error rates, $\eta=\lbrace 0,0.1,0.2 \rbrace$. For each error rate, the number of epoch is also varied among 1, $n$, and $n^2$. The number of samples needed increases (decreases) as the error rate (number of epoch) increases.}
\end{figure}

In the R-LPN algorithm, the time complexity is again dominated by state preparation, for which the number of steps increases at least linearly with the number of samples as mentioned in the previous section. The computation time for calculating the Hamming distance between $M$ guessed parity bit strings and the actual parity bit string is $O(nM^2)$. Since these computation times depend on $M$, we focus on estimating the sample cost. Using the above equation, the number of samples required for achieving $\mathbbm{P}(\tilde{s}_M=s)>2/3$, denoted by $M_{2/3}$, can be calculated numerically. Figure~\ref{fig:numerical} shows $M_{2/3}$ as a function of $n$ for several values of the error probability, $\eta=\lbrace 0,0.1,0.2\rbrace$. For each error rate, the number of epoch is given as 1, $n$, and $n^2$. When $n=40$, there are $2^{40}\approx 10^{12}$ possibilities for $s$. But even in the presence of a relatively high error probability of $20\%$, having only about $10^4$ examples suffices to solve the problem. Figure~\ref{fig:numerical} also suggests that the number of samples can be further reduced by increasing epoch.

\begin{figure}[t]
\centering
\includegraphics[width=0.7\columnwidth]{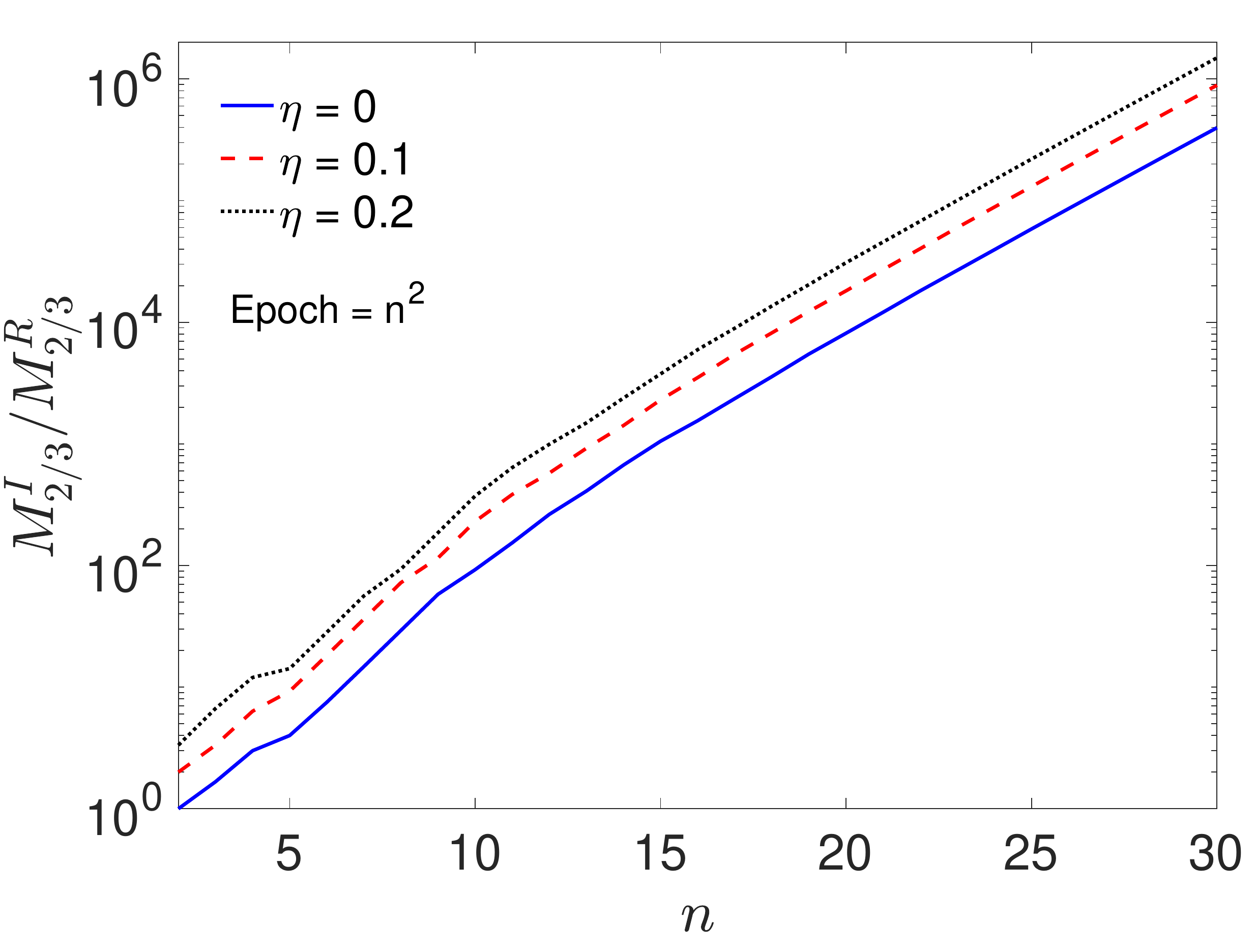}
\caption{\label{fig:comparison}Ratio between the numbers of samples required for achieving $\mathbbm{P}(\tilde{s}_M=s)>2/3$ in I-LPN, denoted by $M^I_{2/3}$, and in R-LPN, denoted by $M^R_{2/3}$ as a function of the length of the hidden bit string, $n$, for various error rates, $\eta=\lbrace 0,0.1,0.2 \rbrace$. For all calculations, the number of epoch is $n^2$.}
\end{figure}

We compare $M_{2/3}$ of I-LPN and R-LPN as a function of $n$ for several values of the error probability, $\eta=\lbrace 0,0.1,0.2\rbrace$, in Fig.~\ref{fig:comparison}. In this comparison, the number of epoch is $n^2$. The result shows that R-LPN reduces the sample complexity by several orders of magnitude for when $n$ is only 15 or so, and this improvement continues to increase as $n$ increases. When $n$ is about 15 to 30, the curves qualitatively suggests that R-LPN enhances the sample complexity exponentially in $n$. However, our analysis does not provide a definitive conclusion about the rate of improvement in the asymptotic limit.

\subsection{Simulation}
\label{sec:4.3}
We use simulations to verify the performance of the R-LPN algorithm, and to compare to known classical methods that are listed in Tab.~\ref{tab:1}. Each iteration starts with the quantum state of the form shown in step 7 of Alg.~\ref{alg:2}, using classical data that are provided uniformly at random. The simulation then proceeds by following the subsequent steps in Alg.~\ref{alg:2}. For a fixed value of $s$, all simulations are repeated 200 times to average over the set of examples drawn uniformly at random.

\subsubsection{Data filtering}
All simulations used an additional pre-processing step, which we refer to as \textit{data filtering}, as an optional attempt to filter out erroneous examples and improve the success probability. Data filtering counts the number of occurrence of example pairs, $(x_j,f'(x_j))$, denoted by $o_j$. Then, an example with a label $k$ that appears less than some fraction of $\max_j(o_j)$ times (i.e., $o_k<w \max_j(o_j)$/2) is discard. An intuitive motivation behind this procedure is that since examples are randomly drawn from a uniform distribution, the same data can be drawn multiple times and erroneous samples are less frequently queried than correct samples for $\eta<1/2$. The optimal choice of the filtering coefficient, $w$, depends on the error rate. For example, when $\eta=0$, such data filtering is not desired since all data are error-free. However, in our simulations, we assume that $\eta$ is unknown, and we used $w=0.4\sqrt{d_H(g^{M}_{1:M},f'_{1:M})}+0.8$, which was empirically found to perform well in overall for various values of $\eta$. This choice means that the level of data filtering increases as the number of samples, and hence the success probability, increases.

\subsubsection{Results}
\begin{figure}[t]
\centering
\includegraphics[width=1\columnwidth]{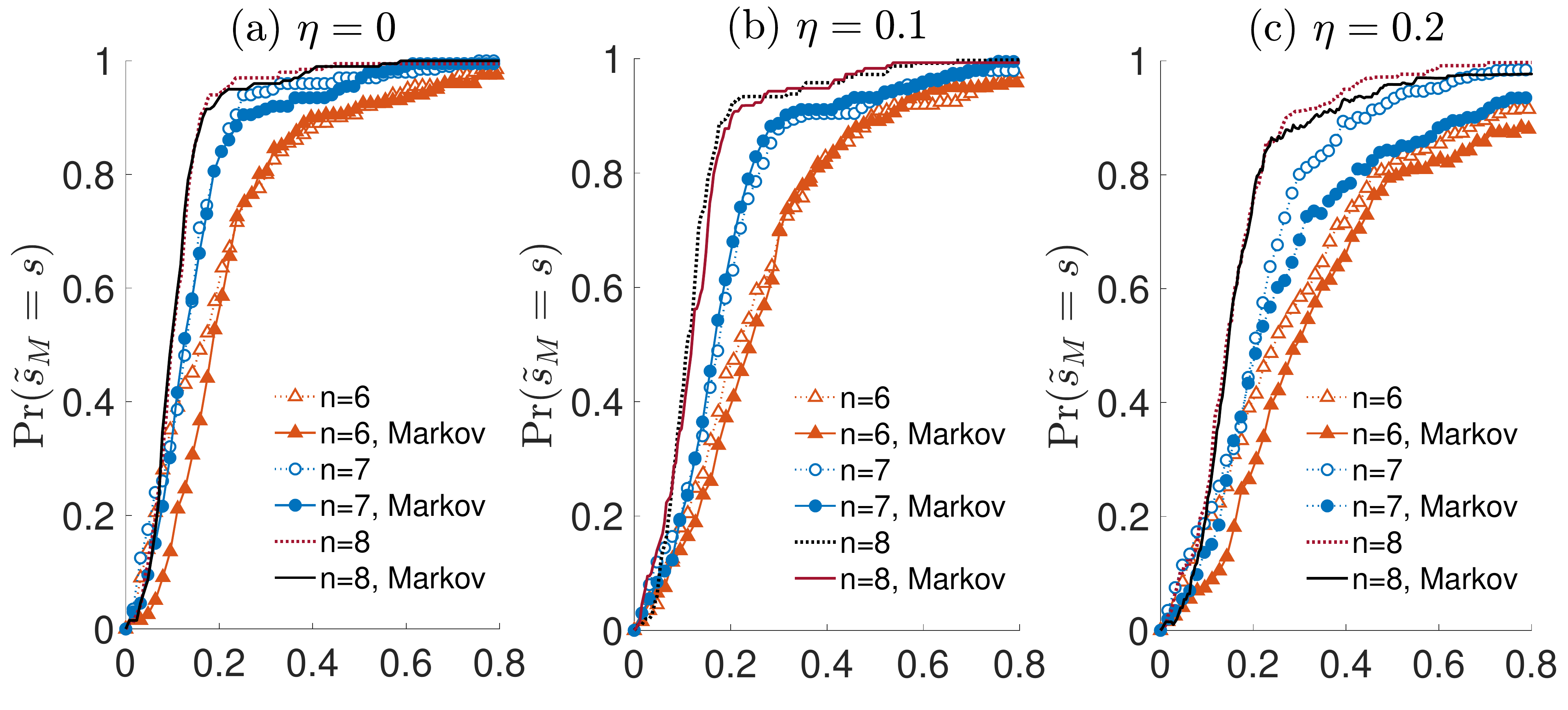}
\caption{\label{fig:simulation_MDP}The plots show the probability to find a hidden bit string $s$ in R-LPN algorithms as a function of the normalized number of samples, $M/2^n$. Dotted lines represent the simulation results of the R-LPN algorithm described in Alg.~\ref{alg:2}, and solid lines represent the simulation results of the modified R-LPN algorithm that keeps only the parity guess function from the previous round, and labelled as \textit{Markov} in the legend. Simulations are performed for (a) $\eta=0$, (b) $\eta=0.1$, and (c) $\eta=0.2$, and for $n=6$, $7$, and $8$. The number of epoch is 30 in all simulations in this figure.}
\end{figure}

We first simulate an R-LPN algorithm with a slight modification, which is intended to save the memory and time cost for storing all $M$ sets of guessed parity bits to calculate their Hamming distances with respect to the real parity bits. Namely, only the guessed parity bits from the previous iteration is kept in the memory, and the greedy algorithm updates the parity guess function for the next round by only comparing the rewards given by the present and the previous guesses. We compare the performance of this modified algorithm to the originally proposed R-LPN by analyzing the success probability with respect to the number of samples via simulations. The simulation results are depicted in Fig.~\ref{fig:simulation_MDP} for $n=6$, $7$, and $8$ as an example, and show that the modified R-LPN does not introduce any considerable change in the sample complexity, especially around the region for $\mathbb{P}(\tilde{s}_M=s)=2/3$. Since the modified R-LPN algorithm uses only the current state for making an action in the subsequent round, the simulation results of this algorithm are denoted by \textit{Markov} in the figure legend.
Hereinafter, all simulations use the modified R-LPN, since it performs similarly to the original version in terms of the sample complexity while the memory and time cost for calculating the rewards does not increase with $M$.

\begin{figure}[t]
\centering
\includegraphics[width=1\columnwidth]{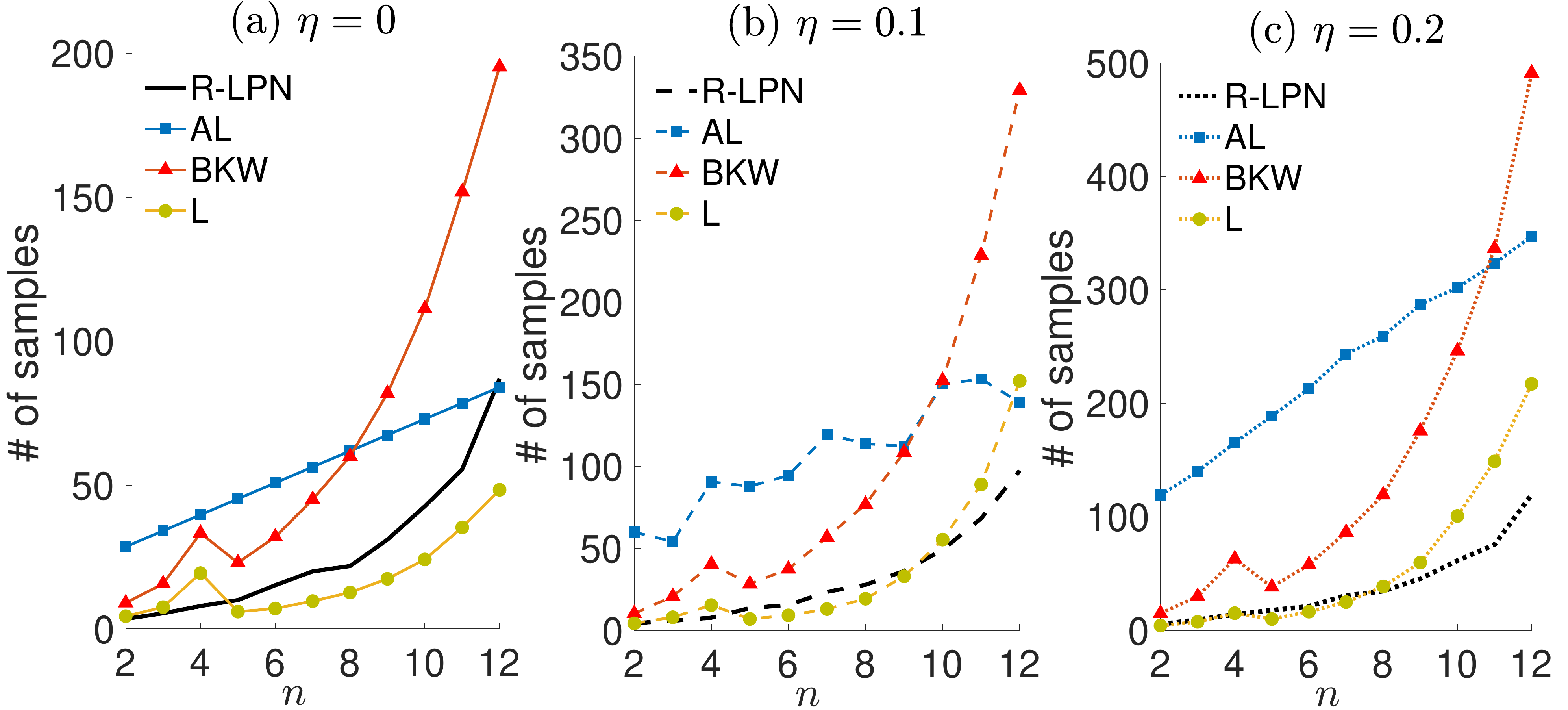}
\caption{\label{fig:simulation_n}Simulation results for the number of samples required for succeeding an LPN algorithm as a function of $n$ for (a) $\eta=0$, (b) $\eta=0.1$, and (c) $\eta=0.2$. Curves without symbols represent the simulation results for the R-LPN algorithm of this work. Simulation results of known classical methods listed in Table~\ref{tab:1} are also plotted and indicated by squares for AL, triangles for BKW, and circles for L. The number of epoch is $n^2$ in all simulations in this figure.}
\end{figure}

Figure~\ref{fig:simulation_n} shows the simulation results of the number of training samples required for succeeding various LPN algorithms as a function of $n$ for several values of the error rate, $\eta$. The number of epoch is $n^2$ in all simulations in this figure. The simulation results show that the R-LPN algorithm performs better than the known classical algorithms, AL~\cite{Angluin1988} and BKW~\cite{Blum2003} (see Tab.~\ref{tab:1}), in the regime of $n\le 12$. In this regime, the algorithm by Lyubashevsky (denoted by L)~\cite{Lyubashevsky2005} consumes the least amount of samples among the classical methods. When $\eta=0$, R-LPN appears to perform slightly worse than L. However, R-LPN becomes advantageous for learning in the presence of the noise, especially as $n$ increases. As summarized in Tab.~\ref{tab:1}, the run time of L is subexponentially greater than its sample complexity. Moreover, the run time of BKW is comparable to its sample complexity and the run time of AL increases exponentially with respect to $n$. However, the run time of R-LPN is expected to be comparable to its sample complexity. Thus, we expect the R-LPN algorithm to provide faster learning compared to the classical algorithms.

\begin{figure}[t]
\centering
\includegraphics[width=1\columnwidth]{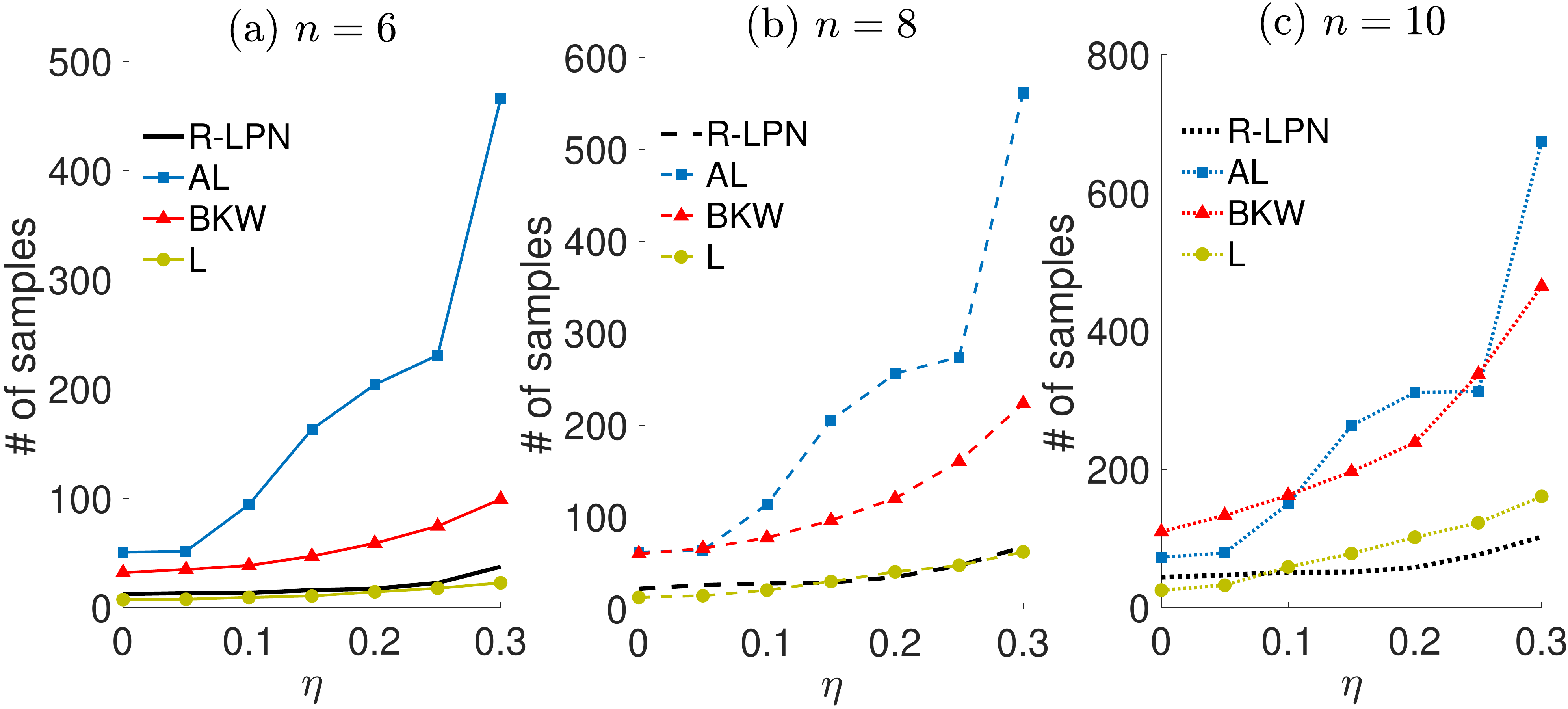}
\caption{\label{fig:simulation_p}Simulation results for the number of samples required for succeeding an LPN algorithm as a function of $\eta$ for (a) $n=6$, (b) $n=8$, and (c) $n=10$. Curves without symbols represent the simulation results for the R-LPN algorithm. Simulation results of known classical methods listed in Tab.~\ref{tab:1} are also plotted and indicated by squares for AL, triangles for BKW, and circles for L. The number of epoch is $n^2$ in all simulations in this figure.}
\end{figure}

The R-LPN algorithm is also more resilient to noise as demonstrated in Fig.~\ref{fig:simulation_p}. The simulation results show that the number of samples needed for succeeding aforementioned LPN algorithms as a function of the error rate, $\eta$, for various $n$. The number of epoch in all simulations in this figure is $n^2$. The R-LPN algorithm requires less number of samples than AL and BKW in all instances in simulations. R-LPN and L algorithms perform similarly for small $n$, but one can see that R-LPN prevails as $n$ increases to 10. From this trend, we speculate that the advantage of R-LPN over the classical algorithms in terms of the robustness to classical noise can become greater as the problem size increases.

Note that by increasing the number of epoch, the number of required samples can be further reduced, at the cost of increasing the run time.

\subsubsection{$\epsilon$-greedy Algorithm}
We also tested an $\epsilon$-greedy algorithm as the policy for making an action via simulation with $n\le 12$. Here, $\tilde{s}_M$ that maximizes the reward is used to guess the quantum oracle with a probability of $1-\epsilon$, and a randomly guessed $n$-bit string is chosen as $\tilde{s}_M$ with a probability of $\epsilon$. The simulation shows that the $\epsilon$-greedy algorithm does not provide any noticeable improvement.

\section{Robustness to Pauli Errors}
\label{sec:5}
Without loss of generality, we assume that the eigenstates of the $\sigma_z$ operator constitute the computational basis. Then since the R-LPN algorithm performs the measurement in the $\sigma_z$ basis, it is not affected by any error that effectively appears as unwanted phase rotations at the end of the quantum circuit.

According to Ref.~\cite{LPNTheory}, when independent bit-flip errors occur on the final state with a probability $\eta_x$, a bit-wise majority vote on $k$ post-selected bit strings gives an estimate $\hat{s}$ such that the error can be bounded as $\mathbbm{P}(\hat{s}\neq s)<4n \exp (-kO(\text{poly}(1/2-\eta_x)))$. When the R-LPN algorithm is completed, it outputs the same final state as in Ref.~\cite{LPNTheory} with high probability. Hence the above result can be directly applied to our algorithm for bit-flip errors on the final state. In this case, the algorithm needs to perform the bit-wise majority vote at each cycle of querying a sample, increasing the total run time. Therefore, quantum noise in the R-LPN algorithm that effectively accumulates on the final state as bit-flip errors with an error rate of $\eta_x<1/2$ only increases the time complexity by a factor of $O(\log(n)\text{poly}(1/(1/2-\eta_x)))$, while the sample complexity remains the same.

\section{Conclusion}
\label{sec:6}
The quantum speed-up in the learning parity with noise problem diminishes in the absence of the quantum oracle that provides a quantum state that encodes all possible examples in superposition upon a query. We developed a quantum-classical hybrid algorithm for solving the LPN problem with classical examples. The LPN problem is particularly challenging as it requires the exact solution to be found. Our work demonstrates that the concept of variational quantum algorithms can be extended for solving such problems. The reinforcement learning significantly reduces both the sample and the time cost of the quantum LPN algorithm in the absence of the quantum oracle. Simulations in the regime of small problem size, i.e., $n \le 12$, show that our algorithm performs comparably or better than the classical algorithm that performs the best in this regime in terms of the sample complexity. The sample cost can be further reduced by increasing the number of epoch, at the cost of increasing the run time. In terms of the vulnerability to noise, our algorithm performs better than classical algorithms in this regime. Furthermore, time complexity can be reduced substantially, if an efficient procedure for updating the quantum state is available.

The ability to utilize quantum mechanical properties to enhance existing classical methods for learning from classical data is a significant milestone towards practical quantum learning. In particular, whether the known advantage of oracle-based quantum algorithms can be retained in the absence of the quantum oracle is an interesting open problem. We showed that for the LPN problem, quantum advantage can be achieved with the integration of classical reinforcement learning. 

Our results motivate future works to employ similar strategies to known oracle-based quantum algorithms in order to extend their applicability to classical data. For example, extending the idea of the quantum-classical reinforcement learning to the learning with errors problem~\cite{PhysRevA.99.032314} would be an interesting future work. This work only considered classical noise and a simple quantum noise model, and detailed studies on the effects of actual experimental quantum errors remains as the future work.

\begin{acknowledgments}
We thank Suhwang Jeong and Jeongseok Ha for stimulating discussions. This work was supported by the Ministry of Science and ICT, Korea, under an ITRC Program, IITP-2019-2018-0-01402, and by National Research Foundation of Korea (NRF-2019R1I1A1A01050161).
\end{acknowledgments}

\section*{Conflict of interest}
The authors declare that they have no conflict of interest.

\end{document}